\documentclass[12pt]{article}    
\usepackage{latexsym}
\usepackage{amsmath}
\usepackage{amssymb}
\usepackage{proof}
\usepackage{fullpage}

\begin{document}
\newcommand{\oto}{\mathrel{\relbar\joinrel\relcirc}}
\newcommand{\bang}{\mbox{\bf    \,!\,}}    
\newcommand{\sM}{{\sf   M}}
\newcommand{\DVS}{{\sf      DVS}}     
\newcommand{\CBS}{{\sf CBS}}
\newcommand{\Born}{{\sf Born}}    
\newcommand{\tCBS}{{\sf tCBS}}
\newcommand{\stCBS}{{\sf stCBS}}    
\newcommand{\Con}{{\sf Con}}
\newcommand{\LCS}{{\sf LCS}}
\newcommand{\lspace}{\mbox{$\ell^\infty$-{\sf Space}}}
\newcommand{\cspace}{\mbox{${\cal C}^\infty$-{\sf Space}}}
\newcommand{\lvec}{\mbox{$\ell^\infty$-{\sf Vec}}}
\newcommand{\cvec}{\mbox{${\cal C}^\infty$-{\sf Vec}}}
\newtheorem{thm}{Theorem}[section] 
\newtheorem{prop}[thm]{Proposition}
\newtheorem{lem}[thm]{Lemma}           
\newtheorem{cor}[thm]{Corollary}
\newtheorem{defn}[thm]{Definition}   
\newtheorem{conj}[thm]{Conjecture}
\newtheorem{expl}[thm]{Example}           
\newtheorem{rem}[thm]{Remark}
\newtheorem{alg}[thm]{Algorithm}        
\newcommand{\rarr}{\rightarrow}
\newcommand{\Nat}{\ensuremath{\mathbb{N}}}
\newcommand{\R}{\ensuremath{\mathbb{R}}}
\newcommand{\stt}[1]{\stackrel{#1}{\longrightarrow}}
\newcommand{\relcirc}{\mathrel{\circ}}
\newcommand{\limp}{\mathrel{\relbar\joinrel\relcirc}} %Linear hom arrow
\newcommand{\ofrom}{\mathrel{\relcirc\joinrel\relbar}} %Lin hom back-arrow
\renewcommand{\o}{\circ}                       
\newcommand{\oa}{\oplus}
\newcommand{\ox}{\otimes}               
\newcommand{\bigox}{\bigotimes}
\newcommand{\Rarr}{\Rightarrow}
\newcommand{\Larr}{\Leftarrow}
\newcommand{\bigoa}{\bigoplus}
\newcommand{\tadj}{\dashv} % in-text adjunction symbol
\newcommand{\ent}{\vdash}
\newcommand{\alt}{\mathrel{|}}
\newcommand{\imp}{\Rightarrow}
\newcommand{\deq}{\stackrel{\rm{def}}{=}}
\newcommand{\reals}{\mathbb{R}}
\newcommand{\nats}{\mathbb{N}}
\newcommand{\y}{{\sf Y}}
\newcommand{\cC}{{\cal C}}
\newcommand{\cM}{{\cal M}}
\newcommand{\cV}{{\cal V}}
\newcommand{\cE}{{\cal E}}
\newcommand{\const}[1]{\mathrm{const}_{#1}}

 \def\pushright#1{{%              set up
    \parfillskip=0pt            % so \par doesnt push \square to left
    \widowpenalty=10000         % so we dont break the page before \square
    \displaywidowpenalty=10000  % ditto
    \finalhyphendemerits=0      % TeXbook exercise 14.32
   %
   %                 horizontal
    \leavevmode                 % \nobreak means lines not pages
    \unskip                     % remove previous space or glue
    \nobreak                    % don't break lines
    \hfil                       % ragged right if we spill over
    \penalty50                  % discouragement to do so
    \hskip.2em                  % ensure some space
    \null                       % anchor following \hfill
    \hfill                      % push \square to right
    {#1}                        % the end-of-proof mark (or whatever)
   %
   %                   vertical
    \par}}                      % build paragraph

 % prefer proofs with statements, also space after
 \def\qed{\pushright{$\Box$}\penalty-700 \smallskip}

%% =====================================================================
% Robin's prf environment - modified with PT's \qed above -
%
% For laying out a  proof: \begin{prf}{} ... \end{prf} The argument is
% in case the proof is a continuation as in:
% \begin{prf}{\ref{four-color} continued} ...  \end{prf} if you do not
% use the argument be careful to use the brackets
%
% As originally done by Robin:
% \newenvironment{prf}[1]{\begin{trivlist} \item[{\bf ~Proof}#1.]}%
% {\begin{flushright} $ \Box $  \end{flushright}\end{trivlist}}
% As modified by RAGS:
\newenvironment{prf}[1]{\begin{trivlist} \item[{\bf ~Proof}#1.]}%
{\qed\end{trivlist}}

\title{A convenient differential category}
 \author{
Richard Blute\thanks{
 Department of Mathematics, University of Ottawa, 
 Ottawa, Ontario, K1N 6N5, CANADA 
 {\tt rblute@uottawa.ca}. 
\quad Research supported by an NSERC Discovery Grant. }
 \and
 Thomas Ehrhard\thanks{Laboratoire PPS,
Universit\'e Paris Diderot - Paris 7,
Case 7014, 
75205 Paris Cedex 13, FRANCE, {\tt thomas.ehrhard@pps.jussieu.fr}.
\quad Research supported by the french ANR project Choco (ANR-07-BLAN-0324).}
 \and
 Christine  Tasson\thanks{CEA LIST, Laboratory  for the  Modelling and
   Analysis  of   Interacting  Systems,  Point  Courrier   94,  91  191
   Gif-sur-Yvette, FRANCE, {\tt christine.tasson@cea.fr}.
\quad Research supported by the french ANR project Choco (ANR-07-BLAN-0324).}
}

\maketitle

\begin{abstract}
  In  this  paper,  we  show  that the  category  of  Mackey-complete,
  separated,  topological   convex  bornological  vector   spaces  and
  bornological linear  maps is  a differential category.   Such spaces
  were introduced  by Fr\"olicher and  Kriegl, where they  were called
  {\it convenient vector spaces}.

  While  much   of  the   structure  necessary  to   demonstrate  this
  observation is  already contained in Fr\"olicher  and Kriegl's book,
  we  here give  a new  interpretation of  the category  of convenient
  vector spaces as  a model of the {\it  differential linear logic} of
  Ehrhard and Regnier.

  Rather  than base our  proof on  the abstract  categorical structure
  presented  by Fr\"olicher  and Kriegl,  we  prefer to  focus on  the
  bornological  structure  of convenient  vector  spaces.  We  believe
  bornological  structures will  ultimately  yield a  wide variety  of
  models of differential logics.
\end{abstract}
 
\section{Introduction}

The  {\it differential $\lambda$-calculus}  was introduced  by Ehrhard
and Regnier~\cite{ER1,ER2} in order to describe the differentiation of
higher  order functionals  from  a syntactic  or logical  perspective.
There  are models  of this  calculus~\cite{Ehr1,Ehr2}  with sufficient
analytical  structure to  demonstrate that  the formalism  does indeed
capture differentiation.  But there were no  models directly connected
to differential geometry, which  is of course where differentiation is
of the  highest significance.   The purpose of  this paper is  to show
that convenient  spaces (introduced below)  constitute a model  of the
differential  linear logic,  the  logical system  associated with  the
differential $\lambda$-calculus. {\it Differential linear logic} is an
extension  of  linear  logic   \cite{Gir}  to  include  an  additional
inference rule  to capture differentiation.   Convenient vector spaces
have been suggested  as an excellent framework for  the foundations of
analysis on infinite-dimensional manifolds. See \cite{KM}.

The  question  of how  to  differentiate  functions  into and  out  of
function  spaces  has  a   significant  history.   For  instance,  the
importance of  such structures is fundamental in  the classical theory
of  {\it variational  calculus},  see e.g.~\cite{GF}.   It  is also  a
notoriously difficult  question.  This can be seen  by considering the
category of smooth manifolds  and smooth functions between them. While
products evidently exist in this category, there is no way to make the
set of functions between two  manifolds into a manifold.  This is most
concisely expressed by saying that the category of smooth manifolds is
not cartesian  closed. Thus,  category theory provides  a particularly
appropriate framework for the  analysis of function spaces through the
notion of cartesian closed category.

Furthermore,  in the  categorical  approach to  modelling logics,  one
typically starts  with a logic  presented as a sequent  calculus.  One
then arranges  equivalence classes of derivations into  a category. If
the equivalence relation is chosen wisely, the resulting category will
be   a    free   category   with   structure.     For   example,   the
conjunction-implication  fragment of  intuitionistic logic  yields the
free  cartesian  closed category;  the  times-implication fragment  of
intuitionistic linear logic yields  the free symmetric monoidal closed
category.
In  both these  cases, the  implication  connective is  modelled as  a
function space, i.e.  the right adjoint to product.  So any attempt to
model  the {\it  differential  linear logic}  of  Ehrhard and  Regnier
should be a  category whose morphisms are smooth  maps for some notion
of smoothness.  Then, to  model logical implication, the category must
also be closed.

More precisely, a  significant question raised by the  work of Ehrhard
and Regnier  is to  write down the  appropriate notion  of categorical
model  of differential linear  logic.  This  was undertaken  by Blute,
Cockett and Seely  in \cite{BCS}. There a notion  of {\it differential
  category} is defined  and several examples are given  in addition to
the usual one made of the syntax of the logic.
In a  followup \cite{BCS2}, the  coKleisli category of  a differential
category  is considered  directly  and the  notion  of {\it  cartesian
  differential  category} is  introduced.  Notice  that the  notion of
differential  category is  related  with the  more  classic theory  of
K\"ahler differentials, as shown in \cite{BCPS}, where differentiation
is   considered  via   a  universal   property  similar   to  K\"ahler
differentiation  in  commutative  algebra \cite{Hart,Mats}.   K\"ahler
categories  are introduced  as symmetric  monoidal categories  with an
algebra modality and a  universal derivation.  It is then demonstrated
that every codifferential category with a minor structural property is
K\"ahler.  Otherwise, in~\cite{BEM} the  notion of a {\it differential
  $\lambda$-category}   is  introduced   as  extension   of  cartesian
differential categories  taking account closed  structure. It provides
appropriate   axiomatic   structure    to   model   the   differential
$\lambda$-calculus.

\medskip In this paper, we  focus on the category of convenient vector
spaces and  bounded linear maps,  and demonstrate that it  provides an
ideal  framework for  modelling  differential linear  logic. The  idea
behind   this  structure  is   inspired  largely   by  a   theorem  of
Boman~\cite{Bom}, discussed in \cite{Fro,KM,BH,Sta}.
\begin{thm}[Boman, 1967]
  \begin{itemize}
  \item Let $E$ be a Banach space and $c\colon \R\rarr E$. Then $c$ is
    smooth with respect to the normed  structure of $E$ if and only if
    $\ell\circ c\colon \R\rarr\R$ is smooth in the usual sense for all
    linear, continuous maps $\ell\colon E\rarr\R$.
  \item Let $E$  and $F$ be Banach spaces and $f\colon  E\rarr F$ be a
    function.   Then $f$  is smooth  if and  only if  it  takes smooth
    curves on $E$ to smooth curves in $F$.
  \end{itemize}
\end{thm}
This remarkable theorem suggests  the possibility of defining a notion
of smooth map between topological (or bornological) spaces without any
notion of norm. One needs only to define a reasonable notion of smooth
curve  into a  space  and then  define  general smoothness  to be  the
preservation of smooth curves.

The two monographs \cite{Fro,KM}  as well as numerous papers indicates
how  successful  this  idea  is.  In the  more  abstract  approach  to
convenient  vector spaces, one  considers the  monoid $\cM$  of smooth
maps from  the reals to  the reals. Roughly,  an arbitrary set  $X$ is
equipped with  an $\cM$-{\it  structure} if equipped  with a  class of
functions  $f\colon   \R\rarr  X$,  called   a  set  of   {\it  smooth
  curves}. (This set must satisfy  a closure condition.)  We call such
a structure  an $\cM$-{\it space}.  Then a {\it smooth  functional} on
$X$ is  a function  $g\colon X\rarr \R$  such that its  composite with
every smooth curve is in  $\cM$.  This approach to defining smoothness
also appears in \cite{Lawv}.
A  {\it smooth  function} between  $\cM$-spaces is  a  function taking
smooth  curves to smooth  curves.  Equivalently,  precomposition takes
smooth functionals to smooth functionals.
If one further requires that the set $X$ be a (real) vector space, and
that   the   vector  space   operations   are   compatible  with   the
$\cM$-structure,  we   have  the  notion  of  a   {\it  smooth  vector
  space}. Finally, adding a separation axiom and a bornological notion
of  completeness called  {\it  Mackey completeness},  one  has a  {\it
  convenient vector space}.

While this  abstract approach  to describing convenient  vector spaces
has  its advantages,  there is  another equivalent  description, which
uses ideas  from functional analysis.  There is an  adjunction between
convex  bornological vector  spaces and  locally convex  vector spaces
(definitions  and constructions  explained  below). Convenient  vector
spaces can  be seen as  the fixed points  of the composite of  the two
functors.  This allows  one  to use  classical  tools from  functional
analysis  in their  consideration.  It also  suggests the  appropriate
notion of linear  map. It turns out that the  notion of bounded linear
maps, when  a convenient  vector space is  viewed as  bornological, is
equivalent  to the  notion of  continuous linear  map, when  viewed as
locally convex.

The category  of convenient vector spaces and  bounded (or continuous)
linear maps  has a  number of remarkable  properties. It  is symmetric
monoidal closed,  complete and cocomplete. But  most significantly, it
is equipped with a comonad, for which the resulting coKleisli category
is the category of smooth maps, in the sense defined above.
This is of course reminiscent  of the structure of linear logic, which
provides a  decomposition of  intuitionistic logical implication  as a
linear  implication  composed  with   a  comonad.   The  classic  (and
motivating)  example is  that  the category  of  coherence spaces  and
stable maps can  be obtained as the coKleisli category  for a monad on
the category of coherence spaces and linear maps, see \cite{GLT}.
 
\medskip 
 After  describing  the  category of  convenient  spaces,  we
demonstrate that  it is  a model of  intuitionistic linear  logic, and
that  the  coKleisli  category  corresponding  to  the  model  of  the
exponential modality  is the category  of smooth maps. We  construct a
differential operator on  smooth maps, and show that it  is a model of
the  differential inference  rule  of differential  linear logic.  The
concrete approach taken here makes it much more convenient\footnote{We
  promise that  this is the only  time we make this  pun.} to describe
differentiation.

\medskip 
We  note that  essentially all of  the structure  we describe
here     can    be     found    scattered     in     the    literature
\cite{Fro,Hog,Jar,KM,Treves}, but we  think that presenting everything
from  the  bornological  perspective  sheds  new  light  on  both  the
categorical and logical  structures. For most of the  results, we give
at least sketches of proofs so  that the paper is as self-contained as
possible. We  also give  detailed references, so  that the  reader who
wishes  can  explore further.  We  hope  this  paper demonstrates  the
remarkable  nature of  this  subject initiated  by Fr\"olicher,  which
predates not only differential linear logic, but linear logic, itself.

\bigskip

\noindent {\bf Acknowledgements}: The first author would like to thank
NSERC for its financial support and the second and third authors thank
the French  ANR project  Choco. The authors  would like  to especially
thank Phil Scott for his helpful contributions.

\bigskip

\section{Differential linear logic and differential categories}

We assume  that the  reader is familiar  with the sequent  calculus of
linear   logic    \cite{Gir}.   {\it   Differential    linear   logic}
\cite{ER1,ER2} is an extension  of the traditional sequent calculus to
include an inference rule representing differentiation. The rule is as
follows:
\[\infer{A,\bang A\vdash B}{\bang A\vdash B}\]

The   semantic  interpretation   is   that,  if   an  arrow   $f\colon
\bang\R^n\rarr \R^m$ is  a smooth map from $\R^n$  to $\R^m$, then its
differential   would   be  of   the   form  $df\colon   \bang\R^n\rarr
\R^n\oto\R^m$, where  $\R^n\oto\R^m$ denotes  the linear maps.  So the
differential of  $f$ is a smooth  function that takes a  vector $v$ in
$\R^n$  and  calculates  the  directional  derivative of  $f$  in  the
direction of $v$.  One treats this inference rule as any other sequent
calculus  inference  rule,   and  verifies  cut-elimination  for  this
extension of linear logic.

A stronger sequent calculus is given  by assuming the duals of some of
the usual structural rules of linear logic:

\smallskip

\[\infer[codereliction]{\Gamma\vdash\bang A}{\Gamma\vdash A}\]
\[\infer[cocontraction]{\Gamma,\Delta\vdash\bang   A}{\Gamma\vdash\bang
  A\qquad\Delta\vdash\bang A}\]
\[\infer[coweakening]{\vdash\bang A}{}\]

\smallskip

The   {\it  differential   proof   nets}  of   \cite{ER2}  provide   a
graph-theoretic  syntax for specifying  proofs in  the logic,  and the
differential $\lambda$-calculus  of \cite{ER1} is an  extension of the
usual  simply-typed  $\lambda$-calculus  to  include this  rule.  (One
should   think   of  this   $\lambda$-calculus   as   living  in   the
cartesian-closed coKleisli category of the comonad \bang.)

\smallskip

The  appropriate categorical  structure to  model  differential linear
logic  is the  {\it  differential category},  which  is introduced  in
\cite{BCS}. One begins  with the usual notion of  categorical model of
linear  logic, also  known as  a  {\it Seely  model}. So  we assume  a
symmetric,   monoidal  closed   category.\footnote{The  $*$-autonomous
  structure of  classical linear logic  will not concern us  here.} We
also    assume     the    existence    of     a    comonad,    denoted
$(\bang,\rho,\epsilon)$, satisfying a  standard set of properties. See
\cite{Mel}  for  an excellent  overview  of  the  topic.  One  of  the
fundamental properties,  necessary for modeling  the contraction rule,
is that each object of the form $\bang A$ is canonically equipped with
a symmetric  coalgebra structure. Thus the functor  \bang induces what
\cite{BCS} refers to as a {\it coalgebra modality.}

To model the  remaining differential structure, it suffices  to have a
{\it deriving  transformation}, i.e.  a natural  transformation of the
form:
\[d_A\colon A\ox\bang A\longrightarrow \bang A\]
satisfying  evident  equations.   These  equations correspond  to  the
standard rules of calculus:
\begin{itemize}
\item The derivative of a constant is 0.
\item Leibniz rule.
\item The derivative of a linear function is a constant.
\item Chain rule.
\end{itemize}
Each of  these must  be expressed coalgebraically.   We will  give the
equations for a slightly different theory below.

\smallskip
 As in  the syntax, there is an  equivalent presentation. In
the case  where one has  biproducts, then the functor  \bang\ actually
determines a  {\it bialgebra modality}.  So, in addition to  the usual
coalgebra structure:
\[\Delta\colon \bang A\rarr \bang A\ox\bang A,
\qquad e\colon \bang A\rarr I,\]
we also have compatible algebra structure:
\[\nabla\colon\bang A\ox\bang A\rarr\bang
A,\qquad\nu\colon I\rarr\bang A.\]
The  algebra structure  gives us  the interpretations  of  the logical
rules coweakening and cocontraction.   The only remaining structure is
the map corresponding  to codereliction.  Categorically, codereliction
is expressed as a natural transformation:
\[{\sf coder}_A\colon A\longrightarrow \bang A,\]
satisfying certain  equations which are analogues of  the above listed
equations:
\begin{description}
\item[{[dC.1]}] $\quad{\sf coder};e=0$,
\item[{[dC.2]}]     $\quad{\sf    coder};\Delta={\sf     coder}    \ox
  \nu+\nu\ox{\sf coder}$,
\item[{[dC.3]}] $\quad{\sf coder}; \epsilon =1$,
\item[{[dC.4]}] $\quad({\sf coder}\ox 1);\nabla; \rho=({\sf coder} \ox
  \Delta);((\nabla;{\sf coder})\ox\rho)); \nabla$.
\end{description}

\bigskip

\section{Convenient vector spaces}
In this  section, we present  the category of convenient  spaces. They
can be seen either as  topological or bornological vector spaces, with
the two structures satisfying a compatibility.

Let us first recall the more traditional notion of {\it locally convex
  space}.
\begin{defn}{\em A {\it locally  convex space} is a topological vector
    space such  that $0$ has a  neighborhood basis of  convex sets.  A
    morphism  between  locally  convex  spaces  is  simply  a  linear,
    continuous  map.    We  thus   obtain  a  category   denoted  {\sf
      LCS}.  }
\end{defn}
A  locally  convex space  can  equivalently  be  defined as  having  a
topology determined  by a family  of seminorms. See  \cite{Treves} for
details.

\subsection{Bornology}

For the significance of this  structure and an analysis of convergence
properties,  see  \cite{Hog}.   A  set  is  bornological  if,  roughly
speaking, it is equipped with a notion of boundedness.

\begin{defn}{\em A set $X$ is {\it bornological} if equipped with a
{\it bornology}, i.e. a set of subsets ${\cal B}_X$ called bounded such that
\begin{itemize}
\item All singletons are in ${\cal B}_X$.
\item ${\cal B}_X$ is downward closed with respect to inclusion.
\item ${\cal B}_X$ is closed under finite unions.
\end{itemize}
A map  between bornological spaces  is {\it bornological} if  it takes
bounded sets to  bounded sets. The resulting category  will be denoted
\Born.  }
\end{defn}

\begin{thm} The category \Born\ is cartesian closed.
\end{thm}
\begin{prf}{} ({\bf  Sketch}) The product  bornology is defined  to be
  the   coarsest    bornology   such   that    the   projections   are
  bornological. So a  subset of $X\times Y$ is bounded  if and only if
  its two projections are bornological.

  One then defines $X\Rarr Y$  as the set of bornological functions. A
  subset $B\subseteq  X\Rarr Y$  is bounded if  and only if  $B(A)$ is
  bounded in $Y$, for all $A$ bounded in $X$.
\end{prf}
As this  bornology will  arise in a  number of different  contexts, we
will  denote  $X\Rarr  Y$  by  $\Born(X,Y)$. We  note  that  the  same
construction works for products of arbitrary cardinality.

\begin{defn}\label{born}
{\em A {\it convex bornological vector space} is
a vector space $E$ equipped with a bornology such that
\begin{enumerate}
\item ${\cal B}$ is closed under the convex hull operation.
\item If $B\in{\cal B}$, then $-B\in{\cal B}$ and
$2B\in{\cal B}$.
\end{enumerate}
The last condition ensures that addition and scalar multiplication are
bornological maps,  when the reals  are given the usual  bornology.  A
map of convex bornological vector spaces is a linear map such that the
direct image of a bounded set is bounded.  We thus get a category that
we denote {\sf CBS}.  }\end{defn}

\subsection{Topology vs. bornology}

As described  in~\cite{Fro,Jar,Hog}, the  topology and bornology  of a
convenient vector  space are  related by an  adjunction, which  we now
describe.

Let $E$ be a locally convex  space. Say that $B\subseteq E$ is bounded
if it is absorbed by every neighborhood of 0, that is to say if $U$ is
a  neighborhood of  $0$,  then  there exists  a  positive real  number
$\lambda$ such that $B\subseteq\lambda U$. This is called the {\it von
  Neumann   bornology}  associated   to  $E$.   We  will   denote  the
corresponding convex bornological space by $\beta E$.

On the  other hand, let $E$  be a convex bornological  space. Define a
topology on $E$  by saying that its associated  topology is the finest
locally  convex topology  compatible with  the original  bornology. We
will  denote  $\gamma  E$  the  vector space  $E$  endowed  with  this
topology. More concretely, one says  that the bornivorous disks form a
neighborhood basis at  0.  A {\it disk} is a subset  $A$ which is both
convex and  satisfies that $\lambda  A\subseteq A$, for  all $\lambda$
with $|\lambda|\leq  1$. A  disk $A$ is  said to be  {\it bornivorous}
when for  every bounded subsets $B$  of $E$, there  is $\lambda\neq 0$
such that $\lambda B\subseteq A$.

\begin{thm}
  The functor $\beta\colon {\sf  LCS}\rarr {\sf CBS}$ is right adjoint
  to the functor $\gamma\colon  {\sf CBS}\rarr{\sf LCS}$. Moreover, if
  $E$ is a CBS and $F$ a LCS, then $\LCS(\gamma E,F)=\CBS(E,\beta F)$.
\end{thm}
\begin{prf}{}
  Let $f:E\to F$ be a linear map.

  First, assume  that $f$ is bornological  from $E$ to  $\beta F$. Let
  $V$ be a neighborhood of $0$ in $F$. Let us show that $f^{-1}(V)$ is
  a neighborhood of  $0$ in $\gamma E$, that is  a bornivorous. If $B$
  be a bounded subset of  $E$, then, since $f$ is bornological, $f(B)$
  is bounded in  $\beta F$, that is absorbed  by every neighborhood of
  $0$. In  particular, there  is $\lambda>0$ such  that $f(B)\subseteq
  \lambda  V$.   By  linearity  of  $f$, we  get  $B\subseteq  \lambda
  f^{-1}(V)$.  This  concludes the proof  that $f$ is  continuous from
  $\gamma E$ to $F$.

  Second, assume  that $f$ is continuous  from $\gamma E$  to $F$. Let
  $B$  be a bounded  subset of  $E$ and  let us  prove that  $f(B)$ is
  bounded in $\beta F$. Let $V$ be a neighborhood of $0$ in $F$. Since
  $f$ is continuous,  $f^{-1}(V)$ is a neighborhood of  $0$ in $\gamma
  E$, and so  contains a bornivorous $U$.  Now,  $B$ is bounded, hence
  the existence of $\lambda>0$ such that $B\subseteq \lambda U$. Since
  $f$ is  linear, we get $f(B)\subseteq  \lambda f(U)\subseteq \lambda
  V$.   We conclude  that  $f(B)$ is  a  bounded for  the Von  Neumann
  bornology, and hence $f$ is bornological from $E$ to $\beta F$.

  Notice that all along the proof,  we handled the same $f$, hence the
  equality  in  place  of   the  usual  isomorphism  that  proves  the
  adjunction.
\end{prf}

\begin{defn} {\em A convex bornological space $E$ is {\it topological}
    if  $E=\beta\gamma  E$.  A   locally  convex  space  $E$  is  {\it
      bornological} if $E=\gamma\beta E$. }
\end{defn}
For  example, the set  $\R$ of  reals is  a convex  bornological space
which is topological.

\begin{cor}\label{cor:linear-cont-borno}
  If  $E$  and  $F$  are  two  convex  bornological  spaces  that  are
  topological, then  a linear  map $f:E\to F$  is bornological  if and
  only if $f:\gamma E\to\gamma F$ is continuous.
\end{cor}
\begin{prf}{}
  It results from the equalities:
  \begin{equation*}
    \CBS(E,F)=\CBS(E,\beta\gamma F)=\CBS(\gamma E,\gamma F).
  \end{equation*}
\end{prf}

Let  {\sf tCBS}  denote  the full  subcategory  of topological  convex
bornological vector spaces and  bornological linear maps, and let {\sf
  bLCS} denote the category  of bornological locally convex spaces and
continuous linear maps. We note immediately:
\begin{cor}
  The categories {\sf tCBS} and {\sf bLCS} are isomorphic.
\end{cor}
On a side note, remark that  there is a third equivalent category.  By
using the von Neumann bornology  associated to a locally convex space,
one can discuss bornological  maps between locally convex spaces. Thus
we can consider the category of locally convex spaces and bornological
linear maps. Denote this category by {\sf LCSb}.

\begin{lem}[See Theorem 2.4.3 of~\cite{Fro}]
  If  $V$ and  $W$ are  bornological locally  convex spaces,  then the
  notions  of  bornological  linear  map  and  continuous  linear  map
  coincide. The category {\sf LCSb}  is equivalent to {\sf bLCS} under
  the functorial operation of {\it bornologification}.
\end{lem}
Note that  this is only  an equivalence, since many  different locally
convex spaces can yield the same bornological locally convex space.

\smallskip

In what  follows, we will  repeatedly take advantage of  the following
useful characterization. Indeed it entails that, to specify a tCBS, it
suffices to  specify its dual space,  that is the  set of bornological
linear functionals.
\begin{lem}[See Lemma~2.1.23 of \cite{Fro}]
  \label{dual}
  Let $E$ be a convex bornological space.
  $E$ is topological if and only if:
  \begin{quote}
    A subset is bounded if and only  if it is sent to a bounded subset
    of $\R$ by any bornological linear functional.
  \end{quote}
\end{lem}
\begin{prf}{}{\bf (Sketch)}
  Let us first recall the  Mackey theorem, proved for instance p.76 of
  \cite{Hog}.  Let $E$  be a separated locally convex  space. A subset
  of $E$ is  scalarly bounded with respect to  the topological dual if
  and only if it is bounded for the von Neumann topology. This theorem
  can easily  be generalized to non separated  spaces, see Proposition
  2.1.9 of \cite{Fro}.

  Now, let  $E$ be a CBS.  By Corollary~\ref{cor:linear-cont-borno}, a
  linear  functional  $l:E\to  \R$  is  bornological if  and  only  if
  $l:\gamma E\to \R$ is  continuous. Therefore, the Mackey theorem can
  be expressed as:
  \begin{quote}
    A  subset  of  $E$  is   scalarly  bounded  with  respect  to  the
    bornological linear  functionals if and  only if it is  bounded in
    $\beta\gamma E$.
  \end{quote}
  By  definition of  a tCBS,  $E=\beta\gamma E$,  which  concludes the
  proof.
\end{prf}
Thus tCBS's  are precisely  those for which  the bornology  is maximal
with respect to the family  of bornological seminorms or to the family
of bornological linear functionals.

\medskip This can be made  more precise by considering the category of
{\it dualized vector  spaces}.  These will be very  familiar to linear
logicians in that  they are similar to the {\it  Chu space} and double
gluing constructions \cite{Barr,HS,Tan}.

A dualized vector space is a (discrete) vector space $V$ with a linear
subspace $V'$ of the dual space  $V^*$.  There is an evident notion of
map between dualized vector spaces, giving a category {\sf DVS}.
Given  $(V,V')$ a dualized  vector space,  one can  associate to  it a
bornological vector space by saying  that $U\subseteq V$ is bounded if
and only if it is {\it scalarly bounded}, i.e. $\ell(U)$ is bounded in
the reals, for all $\ell$ in $V'$. It is straightforward to check that
this  in fact determines  a functor  $\sigma\colon{\sf DVS}\rarr\CBS$.
The  previous lemma  states that  the topological  convex bornological
spaces are precisely the image of this functor.
\begin{thm}[See Theorem 2.4.3.(i) of~\cite{Fro}]
  The construction  of the  previous paragraph provides  an adjunction
  between the categories {\sf DVS} and \CBS.
\end{thm}

\subsection{Separation and completion}

The  \tCBS's that  we are  interested  in have  the desirable  further
properties of {\it separation} and {\it completion}. We begin with the
easiest of the two notions.
\begin{defn}{\em  A  bornological space  is  {\it  separated} if  $E'$
    separates points,  that is if $x\neq  0\in E$ then  there is $l\in
    E'$ such that $l(x)\neq 0$. }
\end{defn}
One   can  verify  a   number  of   equivalent  definitions   as  done
in~\cite{Fro}, page 53.  For example,  $E$ is separated if and only if
the singleton $\{0\}$ is the only bounded linear subspace.

We  now  introduce  the notion  of  a  completion  with respect  to  a
bornology.   However,  notice  that  bornological  completeness  is  a
different and weaker notion  than topological completeness, so we give
details.
\begin{defn}{\em
 Let $E$ be a bornological space. A net $(x_\gamma)_{\gamma\in\Gamma}$
is {\it Mackey-Cauchy} if there exists a bounded subset $B$ and a net 
$(\mu_{\gamma,\gamma'})_{\gamma,\gamma'\in\Gamma,\Gamma'}$ of real numbers 
converging to 0 such that 
\[x_\gamma-x_{\gamma'}\in\mu_{\gamma,\gamma'}B.\] }\end{defn}
Contrary to what generally happens  in locally convex spaces, here the
convergence of Cauchy nets is  equivalent to the convergence of Cauchy
sequences,  and both  are equivalent  to a  property of  an associated
Banach space.  Given  any hausdorff, locally convex space  $E$ and any
closed  bounded  disk  $B$,   one  can  associate  a  normed  subspace
$E_B\subseteq E$ by
\[E_B=\bigcup_{n\in {\sf N}}\, nB.\]
This   is   a   normed    space   under   the   Minkowski   functional
$p(x)=\inf\{\lambda>0|x\in\lambda B\}$.
\begin{lem} [See Lemma 2.2 of \cite{KM}]
  Let $E$ be a bornological space. Then the following are equivalent:
  \begin{itemize}
  \item Every Mackey-Cauchy net converges.
  \item Every Mackey-Cauchy sequence converges.
  \item For every closed bounded disk $B$, the space $E_B$ is a Banach
    space.
  \end{itemize}
\end{lem}

\begin{defn}{\em
    \begin{itemize}
    \item A  bornological space is {\it Mackey-complete}  if either of
      the above conditions hold.
    \item  A  {\it  convenient  vector space}  is  a  Mackey-complete,
      separated, topological convex bornological vector space.
    \item The  category of  convenient vector spaces  and bornological
      linear maps is denoted \Con.
    \end{itemize}}
\end{defn}

We  note that Kriegl  and Michor  in \cite{KM}  denote the  concept of
Mackey  completenesss as  {\it $c^\infty$-completeness}  and  define a
convenient  vector  space  as  a  $c^\infty$-complete  locally  convex
space. If one takes the  bornological maps between these as morphisms,
then the result  is an equivalent category. We  note that the category
of  convenient   vector  spaces   is  closed  under   several  crucial
operations. The following is easy to check:
\begin{thm}[See  Theorem  2.6.5,   \cite{Fro},  and  Theorem  2.15  of
  \cite{KM}]\hspace{1cm}

  \begin{itemize}
  \item  Assuming that  $E_j$ is  convenient  for all  $j\in J$,  then
    $\prod_{j\in  J}E_j$ is  convenient  with respect  to the  product
    bornology, with $J$ an arbitrary indexing set.
  \item If $E$ is convenient, then  so is $\Born(X,E)$ where $X$ is an
    arbitrary bornological set.
  \end{itemize}
\end{thm}

The remainder of  the section is devoted to  an extension theorem that
allows one to derive a convenient space from any \tCBS.

Let  us first  denote by  \stCBS\  the full  subcategory of  separated
topological   convex  bornological   spaces.  It   is   reflective  in
\tCBS.  More precisely, if  $E$ is  any object  in \tCBS,  then define
$\omega_1(E)$ to be the image of $E$ under the map
\[E\longrightarrow  \prod_{E'}\reals  \mbox{  defined  by  }  v\mapsto
(f(v))_{f}
\]
One can verify  that the kernel of this map is  the closure of $\{0\}$
(with respect to the topology induced by Mackey-Cauchy convergence).
\begin{lem} [See Lemma 2.5.6 of \cite{Fro}]
  Modding  out by  this kernel  induces  an adjoint  to the  inclusion
  functor
  \[\omega_1\colon\stCBS\longrightarrow\tCBS\]
\end{lem}

Let $E$ be an object in \stCBS. We define its completion $\omega_2(E)$
by again considering the map
\[E\longrightarrow  \prod_{E'}\reals  \mbox{  defined  by  }  v\mapsto
(f(v))_{f}.\]
Since $E$ is  separated, this map is now  an embedding.  $\omega_2(E)$
is the closure with respect to Mackey-convergence of the image of this
embedding. We get the following result:

\begin{lem} [See Lemma 2.6.5 of  \cite{Fro}]
  The functor $\omega_2\colon \stCBS\rarr\Con$  is left adjoint to the
  inclusion functor.
\end{lem}
\begin{prf}{}{({\bf Sketch})} Let $g\colon E\rarr F$ be a bornological
  linear  map from  an stCBS  $E$ to  a convenient  vector  space $F$.
  Define the bornological linear map $\bar g$ of convenient spaces by
  \begin{eqnarray*}
    \bar g\colon \quad\prod_{E'}\reals\quad &\rarr& \prod_{F'}\reals\\
    (\lambda_f)_{f\in E'}&\mapsto& (\lambda_{f\circ g})_{f\in F'}.
  \end{eqnarray*}
  Check that  $\bar g(\omega_2(E))\subseteq \omega_2(F)=F$.   Then the
  extension $\omega_2(g)$  of $g$ to $\omega_2(E)$  is the restriction
  of $\bar g$.
\end{prf}

\begin{thm}\label{thm:adjunction-tCBS-Con}
  Combining the previous two adjunctions, we obtain a functor
  \[\omega\colon \tCBS\rarr\Con\]
  which is left adjoint to the inclusion.
\end{thm}

\section{Monoidal structure}

We wish  now to  show that the  category \tCBS\ is  symmetric monoidal
closed.  It  is the category of {\it  Mackey-complete, separated} tCBS
that we  will be primarily interested  in, but it is  of interest that
the symmetric monoidal closed  structure exists already at this level.
Let $E$ and $F$ be tCBS. Using Lemma \ref{dual}, we define a bornology
on its algebraic tensor product by specifying its dual space. Define
\[(E\ox  F)'=\{h\colon E\ox  F\rarr \reals\  |\  \hat{h}\colon E\times
F\rarr \reals \mbox{ is bornological} \} \]
where  $\hat{h}$ refers  to the  associated  bilinear map,  and to  be
bornological means with respect to  the product bornology. A subset of
$E\times  F$  is  bounded if  and  only  if  its two  projections  are
bounded. Evidently, the tensor unit will be the base field $I=\R$.

Now,  let  $L(E,F)$ denote  the  space  of  bornological linear  maps.
$L(E,F)$ obtains a bornology as  a subset of $\Born(E,F)$. We now wish
to prove that it is a \tCBS.
Recall that the dual space is defined by:
\[ L(E,F)'=\{h\colon L(E,F)\rarr \reals\  |\ \mbox{ If $U$ is bounded,
  then $h(U)$ is bounded} \}. \]
\begin{lem}
  A  subset  $U\subseteq L(E,F)$  is  bounded if  and  only  if it  is
  scalarly bounded with respect to the above dual space.
\end{lem}
\begin{prf}{} $\Rarr$ Trivial.

  \smallskip   $\Larr$   Assume   $U\subseteq  L(E,F)$   is   scalarly
  bounded.  Suppose for contradiction  that $U$  is not  bounded.  So,
  there is  some bounded  subset $A$  of $E$ such  that $U(A)$  is not
  bounded in $F$. Since $F$ is  a tCBS, there is a bornological linear
  function $l:F\to\R$ such that $l(U(A))$ is not bounded in $\R$.

  Now, we  can choose a pair  of sequences $(f_n)_{n\in\nats}\subseteq
  U$     and      $(a_n)_{n\in\nats}\subseteq     A$     such     that
  $(l(f_n(a_n)))_{n\in\nats}$ is not bounded in $\R$.  Without loss of
  generality,  we  can  assume  that  $l(f_n(a_n))\geq  4^n$  for  all
  $n\in\nats$.   Set $u=(\frac  1 {2^n})_{n\in\nats}$  and  note that,
  because it is summable, it defines a bornological linear functional:
  \begin{eqnarray*}
    \langle u,\cdot\ \rangle \colon\ \ell^\infty &\longrightarrow &\R\\
    v&\mapsto &\sum_{n=0}^{\infty}u(n)v(n)
  \end{eqnarray*}
  on the set $\ell^\infty$ of bounded sequences on $\R$.  Consider
  \[D=\{((l\circ f_p)(a_n))_{n\in\nats}\,;\, p\in\nats\}\]
  which is  an unbounded  subset of $\ell^\infty$.   Indeed, a  set of
  sequences is  bounded if and  only if the  union of all  elements is
  bounded  in  the  reals.    Recall  that  $f_p\in  L(E,F)$  for  all
  $p\in\nats$ and $(a_n)_{n\in\nats}$ takes  its values in the bounded
  subset $A$ of $E$.  Further, for all natural number $p$, we have
  \[\langle u, ((l\circ f_p)(a_n))_{n\in\nats}\rangle\geq
  u(p)\,l(f_p(a_p))\geq p,\]
  hence $\langle u,D\rangle$ is not bounded in $\R$.
 
  Now, define the linear function
  \begin{eqnarray*}
    \Psi\colon L(E,F)&\longrightarrow& \ell^\infty\\
    f\quad &\mapsto &(l(f(a_n)))_{n\in\nats}.
  \end{eqnarray*}
  As $l$  is bornological,  it is  easy to check  that $\Psi$  takes a
  bounded set $V$ in $L(E,F)$ to a bounded set in $\ell^\infty$.

  Consider the composite
  \[h\colon       L(E,F)\xrightarrow\Psi\ell^\infty\xrightarrow{\langle
    u,\cdot\,\rangle} \reals.\]  It takes bounded sets  in $L(E,F)$ to
  bounded sets of  reals and hence is in the  dual space $L(E,F)'$. As
  $U$ is  assumed to  be scalarly bounded,  $h(U)$ has to  be bounded.
  However,  it contains  the set  of functions  $f_p$  for $p\in\nats$
  whose image under the map $h$  is $\langle u,D\rangle$ and so is not
  bounded. Thus we have a contradiction.
\end{prf}

It follows  from the cartesian closedness  of \Born\ that  there is an
isomorphism
\[L(E_1;E_2,F)\cong L(E_1,L(E_2,F))\]
where  $L(E_1;E_2,F)$  is the  space  of  multilinear  maps. Now,  the
algebraic   tensor  product,  equipped   with  the   above  bornology,
classifies  multilinear  maps. Therefore,  the  above structure  makes
\tCBS\ a symmetric monoidal closed category.

We finally  lifts the symmetric monoidal closed  structure of \tCBS~to
\Con~by defining the tensor product of convenient vector spaces as the
Mackey  completion of  the tensor  product in  \tCBS. The  result then
follows from two standard observations (Section 3.8 of \cite{Fro}):
\begin{itemize}
\item If $F$ is complete, then so is $L(E,F)$.
\item If $E$ and $F$ are separated, then so is $E\ox F$.
\end{itemize}
Hence, we have proved that:
\begin{thm}
  The category \Con\ is symmetric monoidal closed.
\end{thm}

\section{Smooth maps and differentiation}

\subsection{Smooth curves}

The  notion of  a smooth  curve  into a  locally convex  space $E$  is
straightforward.  One  simply has  a  curve  $c\colon  \R\rarr E$  and
defines its derivative by:
\[ c'(t)=\lim_{s\rarr 0} \frac{c(t+s)-c(t)}{s}.\]
Note that this limit is simply the limit in the underlying topological
space  of $E$.  Then,  we define  a curve  to be  {\it smooth}  if all
iterated derivatives exist.  We denote the set of smooth curves in $E$
by $\cC_E$.

In order  to endow $\cC_E$  with a convenient structure,  we introduce
the notion of \emph{different quotients}  which is the key idea behind
the theory of finite difference methods, described in \cite{Lev}.  Let
$\R^{<i>}\subseteq\R^{i+1}$ consist of  those $i+1$-tuples with no two
elements   equal.   It  inherits   its  bornological   structure  from
$\R^{i+1}$. Given  any function $f\colon\R\rarr  E$ with $E$  a vector
space, we recursively define maps
\[\delta^if\colon\R^{<i>}\rarr E \]
by saying $\delta^0f=f$, and then the prescription:
\[
\delta^if(t_0,t_1,\ldots,t                      _i)=\frac{i}{t_0-t_i}\,
[\,\delta^{i-1}f(t_0,t_1,\ldots,t    _{i-1})-\delta^{i-1}f(t_1,\ldots,t
_i)\,]
\]
For example,
\[
\delta^1f(t_0,t_1)=\frac{1}{t_0-t_1}\,[\,f(t_0)-f(t_1)\,].\]
Notice that the extension of this map along the missing diagonal would
be the  derivative of $f$.   There are similar interpretations  of the
higher-order   formulas.   So   these   difference  formulas   provide
approximations to derivatives.

\begin{lem}[See 1.3.22 of \cite{Fro}]\label{lem:lipschitz} Let $c\colon \R\rarr E$ be a function.
  Then $c$  is a smooth curve if  and only if for  all natural numbers
  $i$, $\delta^ic$ is a bornological map.
\end{lem}
\begin{prf}{} ({\bf Sketch})
  If $c$ is  smooth, we will show that there is  a smooth extension of
  $\delta^ic\colon   \R^{<i>}\rarr    E$   to   $\bar{\delta}^ic\colon
  \R^{i+1}\rarr E$.  This  is done by induction.  The  0'th case is by
  definition, and assuming that $\bar{\delta}^{j}c$ exists, define
  \[\bar{\delta}^{j+1}c\colon \R^{j+2}\rarr E \mbox{ \,\,\,\,\, by
    \,\,\,\,\,  } \bar{\delta}^{j+1}c(t,t',\vec{x})=\int_0^1\partial_t
  \bar{\delta}^{j}c(t+s(t'-t),\vec{x})ds,    \]   where   $\partial_t$
  denotes the partial derivative with respect to $t$.

  Now, since $\bar{\delta}^ic$ is smooth, its first derivative is well
  defined. Therefore, its first difference quotient is a Mackey-Cauchy
  net      at     $0$      from     which      we      deduce     that
  $\delta^{i+1}c=\delta^1(\bar{\delta}^ic)|_{\R^{<i+1>}}$            is
  bornological.

  \smallskip

  On  the  other hand,  suppose  that  for  all natural  numbers  $i$,
  $\delta^ic$ is a  bornological map.  We prove by  induction that for
  all $i$,  the $i$'th derivatives  $c^{(i)}$ is well defined  and its
  difference   quotients  are  bornological.    The  $0$'th   case  is
  straightforward.  Assume the difference  quotients of  $c^{(j)}$ are
  bornological.
  In    particular,   $\delta^2c^{(j)}$    is    bornological,   hence
  $c^{(j+1)}=(c^{(j)})'$ exists as the  limit of the Mackey-Cauchy net
  of    $\delta^1c^{(j)}$.    By    calculation,    one   sees    that
  $\delta^ic^{(j+1)}(\vec{x})$ is a linear combination:
  \[\delta^ic^{(j+1)}(\vec{x})=\frac{1}{i+1}\sum_{k=0}^i
  \delta^{i+1}c^{(j)}(\vec{x},t_i)\]
  for   arbitrary  $t_i$.    Since,  for   all  natural   number  $i$,
  $\delta^{i+1}c^{(j)}$     is    bornological,     we     get    that
  $\delta^ic^{(j+1)}$ is bornological
  too. 
\end{prf}

\begin{thm}\label{thm:Con-smooth}
  Suppose $E$ is convenient. Then:
  \begin{quote}
    If  $c\colon \R\rarr  E$ is  a curve  such that  $\ell\circ  c$ is
    smooth for  every bornological linear map  $\ell\colon F\rarr \R$,
    then $c$ is itself smooth.
  \end{quote}
\end{thm}
\begin{prf}{}
  By the preceding lemma, $c$ is  smooth if and only if $\delta^ic$ is
  bornological, for all  $i$. Since $E$ is a  tCBS, this is equivalent
  to saying that for all $\ell\in E'$, we have $\ell\circ\delta^ic$ is
  bornological       and      since       $\ell$       is      linear,
  $\ell\circ\delta^ic=\delta^i(\ell\circ  c)$.  We  conclude by  using
  again the preceding lemma.
\end{prf}

If $X$ and $Y$ are  bornological sets, recall that $\Born(X,Y)$ denote
the bornological space of bornological  functions from $X$ to $Y$ with
bornology as already described.

By Lemma \ref{lem:lipschitz}, the above described difference quotients
define an infinite family of maps:
\[\delta^i\colon\cC_E\rarr \Born(\R^{<i>},E)\]

\begin{defn}{\em Say  that $U\subseteq\cC_E$  is {\em bounded}  if and
    only  if its  image  $\delta^i(U)$ is  bounded  for every  natural
    number $i$.  }
\end{defn}

\begin{thm}
  This structure makes $\cC_E$ a convenient vector space.
\end{thm}
\begin{prf}{} 
  Let us first  prove that $\cC_E$ is a tCBS. A  subset $U$ of $\cC_E$
  is bounded if and only  if its image under every difference quotient
  $\delta^i$ is  bounded in $\Born(\R^{<i>},E)$. Since $E$  is a tCBS,
  this  is equivalent  to saying  that the  image of  $U$  under every
  difference quotient  is scalarly bounded  in $\Born(\R^{<i>},E)$, by
  Lemma \ref{dual}.  We conclude by noting that  any linear functional
  commutes with the difference quotient.

  Separation       follows      from      the       separation      of
  $\prod_{i=0}^\infty\Born(\R^{<i>},E)$. Let us  prove that $\cC_E$ is
  complete. Let  $(c_n)$ be a Mackey-Cauchy sequence  of smooth curves
  into $E$. By  definition of the bornology of  $\cC_E$, we infer that
  $(\delta^ic_n)$ is  a Mackey-Cauchy sequence  of $\Born(\R^{<i>},E)$
  and hence converges. For $i=0$, we get that $(c_n)$ converges and we
  denote  its limit  by $c_\infty$.  We then  prove by  induction that
  $\delta^ic_\infty$ is  the limit of  $(\delta^ic_n)$ and that  it is
  bornological.  We  conclude  that  $c_\infty$  is  smooth  by  Lemma
  \ref{lem:lipschitz}.
\end{prf}

\subsection{Smooth maps}

We are  then left with the question  of how to define  smoothness of a
function between two locally convex spaces.
A  motivation for  the  following definition  is  Boman's theorem,  as
discussed in the Introduction.

\begin{defn}{\em  A function  $f\colon E\rarr  F$ is  {\em  smooth} if
    $f(\cC_E)\subseteq \cC_F$.   Let $\cC^\infty(E,F)$ denote  the set
    of smooth functions from $E$ to $F$.  }
\end{defn}

We  note the obvious  fact that  $\cC_E=\cC^\infty(\R,E)$, as  seen by
considering  $id\colon \R\rarr  \R$ as  a smooth  curve.  In~\cite{KM}
p.19,  it is  shown  that for  linear  maps, it  is  equivalent to  be
bornological and smooth.   In the sequel, we will  only use the direct
sense that we proof below.
\begin{lem}\label{lem:linear+smooth=bornological}%(\cite{KM}, p.19)
  A bornological linear map between convenient spaces is smooth.
\end{lem}
\begin{prf}{}
  Let $f:E\rarr F$  be a linear bornological map  and $c\colon \R\rarr
  E$  a smooth  curve.  By Lemma  \ref{lem:lipschitz}, $\delta^ic$  is
  bornological for all  $i$. Since $f$ is bornological,  so is $f\circ
  \delta^ic$. Now, $f$ is linear and thus commutes with the difference
  quotients  $f\circ  \delta^ic=\delta^i(f\circ  c)$. Again  by  Lemma
  \ref{lem:lipschitz}, $f\circ c$ is smooth.
\end{prf}

Let $\cC^\infty$  denote the category of convenient  vector spaces and
smooth maps. One of the crucial results of \cite{Fro} and \cite{KM} is
that  $\cC^\infty$  is a  cartesian  closed  category.  In fact,  this
category is the coKleisli category of a model of intuitionistic linear
logic,  from  which   the  above  follows.  But  this   is  hardly  an
enlightening proof!  We first give a convenient vector space structure
on $\cC^\infty(E,F)$.

Now, let $E$ and $F$  be convenient vector spaces. If $c\colon \R\rarr
E$      is     a      smooth     curve,      we     get      a     map
$c^\ast\colon\cC^\infty(E,F)\rarr\cC_F$ by precomposing.

\begin{defn}{\em
  Say that  $U\subseteq\cC^\infty(E,F)$ is bounded if and  only if its
  image $c^\ast(U)$ is bounded for every smooth map $c:E\rarr \R$.}
\end{defn}

The  space  $\cC^\infty(E,F)$  has   a  natural  interpretation  as  a
projective limit:

\begin{lem} [See \cite{KM}, p.  30] The space $\cC^\infty(E,F)$ is the
  projective    limit    of    spaces    $\cC_F$,   one    for    each
  $c\in\cC_E$. Equivalently,  it consists of  the Mackey-closed linear
  subspace of
  \[\cC^\infty(E,F)\subseteq\prod_{c\in \cC_E} \cC_F\]
  consisting  of all  $\{f_c\}_{c\in\cC_E}$  such that  $f_{c\,\circ\,
    g}=f_c\circ g$ for every $g\in\cC^\infty(\R,\R)$.
\end{lem}
\begin{prf}{}
  Let us prove that the following linear function is a bijection:
  \begin{eqnarray*}
    \Phi:\cC^\infty(E,F)&\rarr & \prod_{c\in\cC_E}\cC_F\\
    f \ &\mapsto & \{c^\ast(f)\,;\, c\in\cC_E\}.
  \end{eqnarray*}
  Let $f\neq 0$ be a smooth  function from $E$ to $F$.  There is $x\in
  E$  such that $f(x)\neq  0$. Let  $\const x$  be the  constant curve
  equal to $x$.  Then $\Phi(f)_{\const  x}\neq 0$, hence $\Phi$ is one
  to one.

  Now,   consider  $\{f_c\,;\,   c\in\cC_E\}$  such   that  $f_{c\circ
    g}=f_c\circ g$  for all $g\in\cC^\infty(\R,\R)$.   Set $f:x\mapsto
  f_{\const x}(0)$, then for all $c\in\cC_E$, $c^\ast(f)=f_c$. We have
  shown that $\Phi$ is onto.

  To  conclude, the  bornological structure  of  $\cC^\infty(E,F)$ has
  been   defined  to   be  the   bornological  structure   induced  by
  $\prod_{c\in\cC_E}\cC_F$.
\end{prf}

We have shown that  $\cC^\infty(E,F)$ is equivalent to a Mackey-closed
subspace of a convenient vector space. So we have:
\begin{cor}
  The  above  structure makes  $\cC^\infty(E,F)$  a convenient  vector
  space.
\end{cor}

As another consequence  of the above Lemma, we  get a characterization
of smooth curves in $\cC^\infty(E,F)$:
\begin{cor}
  A  curve  $f:\R\rarr\cC^\infty(E,F)$  is   smooth  if  and  only  if
  $c^\ast(f):\R\rarr F$ is smooth for all smooth curves $c$.
\end{cor}

\smallskip We are now able to describe the cartesian closed structure.
\begin{thm}[See Theorem 3.12 of \cite{KM}]
  The category $\cC^\infty$ is cartesian closed.
\end{thm}
\begin{prf}{}

  We need to show that a  map $f\colon E_1\times E_2\rarr F$ is smooth
  if   and    only   if   its    transpose   $\hat{f}\colon   E_1\rarr
  \cC^\infty(E_2,F)$ is smooth.

  Recall  that, by  definition, $\hat{f}$  is  smooth if  and only  if
  $\hat{f}\circ  c_1\colon \R\rarr \cC^\infty  (E_2,F)$ is  smooth for
  all   $c_1\colon   \R\rarr   E_1$   smooth.    In   turn,   by   the
  characterization of smooth curves into $\cC^\infty(E_2,F)$ described
  above, this holds if and only if
  \[(c_2^*\circ\hat{f}\circ c_1)\colon \R\rarr \cC^\infty (\R,F)\]
  is smooth, for all smooth curves $c_2\colon \R\rarr E_2$.
 
  But  this map  $(c_2^*\circ\hat{f}\circ  c_1)$ is  equal to  $f\circ
  (c_1\times   c_2)$.    Thus  the   question   is   reduced  to   the
  one-dimensional version  of Boman's  theorem, proved for  example on
  page 29 of \cite{KM}.
\end{prf}

As  usual, having  a cartesian  closed category  gives us  an enormous
amount of structure to work with, as will be seen in what follows.

\subsection{Differentiating smooth maps}

If these functions  are genuinely to be thought of  as smooth, then we
should  be able to  differentiate them.   That is  the content  of the
following:

\begin{thm}[See \cite{KM}, p. 33] Let $E$ and $F$ be convenient vector
  spaces.  The differentiation operator
  \[d\colon \cC^\infty(E,F)\rarr \cC^\infty(E,L(E,F))\]
  defined as
  \[df(x)(v)=\lim_{t\rarr 0} \frac{f(x+tv)-f(x)}{t}\]
  is  linear and  bounded. In  particular,  this limit  exists and  is
  linear in the variable $v$.
\end{thm}

\begin{prf}{}
  We have an evident map of the form:
  \begin{eqnarray*}
    \cC^\infty(E,F)\times E\times E&\rarr&\cC^\infty(\R,F)\\
    (f,x,v)\quad&\mapsto&[c\colon s\mapsto f(x+sv)].
  \end{eqnarray*}
  If we denote the corresponding curve in $F$ by $c$, then
  \[df(x)(v)=c'(0).\]
  Moreover, if  $f$, $x$ and  $v$ are smooth curves  respectively into
  $\cC^\infty(E,F)$, and the  two occurrences of $E$, then  the map of
  two  variables $c:(s,t)\mapsto f(t)(x(t)+sv(t))$  is smooth  and its
  derivative  $t\mapsto df(t)(x(t))(v(t))$ with  respect to  the first
  variable $s$  is also smooth  in $t$.  Therefore, the  above defined
  function is smooth.
  By   cartesian   closedness,   this   gives   our   desired   smooth
  function \[d:\cC^\infty(E,F)\rarr \cC^\infty(E,\cC^\infty(E,F)).\]

  It remains to prove that for all $f\in\cC^\infty(E,F)$ and $x\in E$,
  $df(x)$  is  linear.  It  is  the standard  calculus  proof  of  the
  linearity of  differentiation.  To  conclude recall that,  thanks to
  Lemma~\ref{lem:linear+smooth=bornological},   every   linear  smooth
  function is bornological, hence $df(x)\in L(E,F)$.
\end{prf}

\section{Exponential structure}

The most difficult  aspect of linear logic \cite{Gir}  from a semantic
point of  view is the  intepretation of the exponential  fragment.  As
noted previously, one  must have a comonad with a  great deal of extra
structure. Moreover, in Theorem 5.1.1  of \cite{Fro}, it is shown that
in this  setting, the comonad precisely  demonstrates the relationship
between  linear maps  and  smooth  maps which  was  envisioned by  the
differential $\lambda$-calculus.

We begin by noting that if  $E$ is a convenient vector space and $x\in
E$,    there    is    a     canonical    morphism    of    the    form
$\delta_x\colon\cC^\infty(E,\R)\rarr       \R$,       defined       by
$\delta_x(f)=f(x)$.   This   is  of   course  the  {\it   Dirac  delta
  distribution}.
\begin{lem}
  The  Dirac  distribution  $\delta\colon E\rarr\cC^\infty(E,\R)'$  is
  smooth.
\end{lem}
\begin{prf}{}
  First, it is  easy to see that $\delta_x$ is  linear for every $x\in
  E$. Let us check it is  bornological. Let $U$ be a bounded subset of
  $\cC^\infty(E,\R)$, that is $c^\ast(U)$ is bounded in $\R$ for every
  smooth curve $c\in\cC_E$. In particular, $\delta_x=\const x^\ast(U)$
  is bounded, and we are done.

  Now,   let   $c$  and   $f$   be   smooth   curves  into   $E$   and
  $\cC^\infty(E,\R)$     respectively.       The     map     $t\mapsto
  \delta_{c(t)}f(t)=f(t)(c(t))$   is  smooth.    Thus,   by  cartesian
  closedness, $\delta$ is smooth.
\end{prf}

\begin{defn}{\em  The  {\em exponential  modality}  $\bang  E$ is  the
    Mackey-closure      of       the      set      $\delta(E)$      in
    $\cC^\infty(E,\R)'$.  }\end{defn}

We will demonstrate that this determines a comonad on \Con.
\begin{thm}\label{thm:bang-adjunction}
  We have the following canonical adjunction:
  \[\cC^\infty(E,F)\cong L(\bang E,F)\]
\end{thm}
\begin{prf}{}
  We establish the  bijection, leaving the straightforward calculation
  of naturality to the reader.   So let $\varphi\colon \bang E\rarr F$
  be a bornological  linear map.  Define a smooth map  from $E$ to $F$
  by $\hat{\varphi}(e)=\varphi(\delta_e)$.

  Conversely, suppose  $f\colon E\rarr  F$ is a  smooth map.  Define a
  linear     map    $\tilde{f}\colon     \bang     E\rarr    F$     by
  $\tilde{f}(\delta_e)=f(e)$.   Let   us  show  that   $\tilde  f$  is
  bornological.  Let  $U\subseteq \delta(E)$ bounded as  a subspace of
  $\bang E$,  that is  as a subset  of $\cC^\infty(E,\R)'$.  The image
  $\tilde f(U)$ is  equal to $U(\{f\})$ which is  bounded as the image
  of  a singleton  set.  By  Theorem~\ref{thm:adjunction-tCBS-Con}, we
  can then extend $f$ to  the Mackey completion of $\delta(E)$ through
  the functor $\omega$. We get a function $\bar f:\bang E\rarr F$.

  It  is  clear  that  this   determines  a  bijection  and  hence  an
  adjunction.
\end{prf}

We now describe the structure that comes out of this adjunction:
\begin{itemize}
\item  The counit is  the linear  map $\epsilon\colon\bang  E\rarr E$,
  defined by  $\epsilon(\delta_x)=x$, and then  extending linearly and
  applying the adjunction $\omega$.
\item The unit is a smooth map $\iota\colon E\rarr\bang E$, defined by
  $\iota(x)=\delta_x$.
\item  The associated comonad  has comultiplication  $\rho\colon \bang
  E\rarr \bang\bang E$ given by $\rho(\delta_x)=\delta_{\delta_x}$.
\end{itemize}

\begin{lem}[See Lemma 5.2.4 of~\cite{Fro}]
  The fundamental isomorphism is satisfied:
  \[\bang(E\times F)\cong \bang E\ox\bang F\]
\end{lem}
\begin{prf}{}
  The trick, as usual, is  to verify that $\bang(E\times F)$ satisfies
  the universal property of the tensor product.

  First  we   note  that  there   is  a  bilinear   map  $m\colon\bang
  E\times\bang  F\rarr  \bang(E\times F)$.   Consider  the smooth  map
  $\iota_{E\times   F}\colon  E\times  F\rarr\bang(E\times   F)$.   By
  cartesian    closedness,    we   get    a    smooth   map    $E\rarr
  \cC^\infty(F,\bang(E\times  F))$,  which  extends  to a  linear  map
  $\bang     E\rarr\cC^\infty(F,\bang(E\times     F))\cong     L(\bang
  F,\bang(E\times F))$. The transpose is the desired bilinear map.  It
  satisfies $m\circ(\iota_E\times\iota_F)=\iota_{E\times F}$.

  We check that $m$ satisfies the appropriate universality.
  Assume  $f\colon  \bang E\times\bang  F\rarr  G$  is a  bornological
  bilinear   map.    Let   us   show   that  $f$   is   smooth.    Let
  $(c_1,c_2)\colon\R  \rarr  \bang  E\times   \bang  F$  be  a  smooth
  curve. We want to show  that $t\mapsto f(c_1(t),c_2(t))$ is a smooth
  curve  into  $G$.   Thanks  to Theorem~\ref{thm:Con-smooth},  it  is
  sufficient to show that for every linear bornological functional $l$
  over $G$, the real function $l\circ f\circ (c_1,c_2)\colon\R\rarr\R$
  is smooth.

  Now, notice that, from simple calculations, we get
  \[\delta^1(l\circ f\circ (c_1,c_2))=l\circ f\circ
  (\delta^1(c_1),c_2)+ l\circ f\circ (c_1,\delta^1(c_2))\]
  and hence $\delta^1(l\circ  f\circ (c_1,c_2))$ is bornological. More
  generally, every difference quotient of $l\circ f\circ (c_1,c_2)$ is
  bornological.   From Lemma~\ref{lem:lipschitz},  we get  that  it is
  smooth.   Then, in turn,  $f\circ(\iota\times\iota)$ is  smooth.  By
  Theorem~\ref{thm:bang-adjunction},  $f$   lifts  to  a   linear  map
  $\bar{f}\colon  \bang(E\times F)\rarr  G$.  By  definition,  $\bar f
  \circ \delta_{(x_1,x_2)}=f(x_1,x_2)$.
  Hence $f$ factors through $m$ and $\bar f$.

  Therefore, the universal property is satisfied by $\bang(E\times F)$
  which is hence isomorphic to $\bang E\otimes \bang F$.
\end{prf}

We conclude:
\begin{thm}
  The category \Con\ is a model of intuitionistic linear logic.
\end{thm}

It is straightforward to see that this category has finite biproducts,
as required in a differential  category. The bialgebra structure is as
follows:

\begin{itemize}
\item   $\Delta\colon    \bang   A\rarr   \bang    A\ox\bang   A$   is
  $\Delta(\delta_x)=\delta_x\ox\delta_x$, and  then extending linearly
  and using the functor $\omega$ to extend to the completion.
\item $e\colon \bang A\rarr I$ is $e(\delta_x)=1$.
\item   $\nabla\colon    \bang   A\ox\bang   A\rarr    \bang   A$   is
  $\nabla(\delta_x\ox\delta_y)=\delta_{x+y}$.
\item $\nu\colon I\rarr\bang A$ is $\nu(1)=\delta_0$.
\end{itemize}

Thus by results of \cite{ER1,ER2,BCS},  it remains to establish a {\it
  codereliction map} of the form:
\[coder\colon E\rarr \bang E\]

The  map  $\iota$  defined  above  almost  meets  the  requirement  of
codereliction but fails to be linear. Fortunately, its differential at
0 will supply us with the desired map.

We can eventually state our main theorem:
\begin{thm}
  The category  \Con\ is  a differential category,  with codereliction
  given by
  \[coder(v)=d\iota(v\ox\delta_0)=\lim_{t\rarr
    0}\frac{\delta_{tv}-\delta_0}{t}\]
\end{thm}
\begin{prf}{}
  It remains only to verify the equations of the definition. These all
  follow from the fact that  the derivative in this category really is
  just  a  derivative  in  the  usual  sense. We  verify  two  of  the
  equations:

  Consider  [dC.2], which  is the  Leibniz rule.  Using the  fact that
  limits  in  the  tensor  product are  calculated  componentwise,  we
  calculate as follows:
  The lefthand side gives
  \[v\mapsto                                   \Delta\left(\lim_{t\rarr
      0}\frac{\delta_{tv}-\delta_0}{t}\right)=             \lim_{t\rarr
    0}\frac{\Delta(\delta_{tv})-\Delta(\delta_0)}{t}=      \lim_{t\rarr
    0}\frac{\delta_{tv}\ox\delta_{tv}-\delta_0\ox\delta_0}{t}.
  \]
  Thanks to a standard calculation, this is
  \[\lim_{t\rarr 0}\left(\frac{\delta_{tv}-\delta_0}{t}\ox\delta_{tv}+
    \delta_0\ox\frac{\delta_{tv}-\delta_0}{t}\right)
  \]
  which is equal  to the righthand side by exchange  of limit with sum
  and tensor.

  \bigskip

  Now   consider    [dC.3].    Let    $v\in   E$.    We    must   show
  $(coder;\epsilon)v=v$. This comes from the calculation:
  \[v\mapsto\lim_{t\rarr
    0}\frac{\delta_{tv}-\delta_0}{t}\mapsto\epsilon\left(\lim_{t\rarr
      0}\frac{\delta_{tv}-\delta_0}{t}\right)             =\lim_{t\rarr
    0}\frac{\epsilon(\delta_{tv}-\delta_0)}{t}=\lim_{t\rarr
    0}\frac{tv-0}{t}=v.\]
\end{prf}

\section{Conclusion}

Fundamental to understanding the structure of convenient vector spaces
is the  duality between  bornology and topology  in the  definition of
\tCBS. Another  place where there is  such duality is the  notion of a
finiteness space, introduced in \cite{Ehr2}. But there, the duality is
between  bornology   and  the  {\it  linear   topology}  of  Lefschetz
\cite{Lef}. The advantage of the  present setting is that the topology
takes   place  in   the  more   familiar  world   of   locally  convex
spaces.  However, it  remains an  interesting question  to work  out a
similar structure in the Lefschetz setting. This program was initiated
in the thesis of the third author \cite{Tas}.

Evidently,  a  next  fundamental  question  is  the  logical/syntactic
structure  of integration.  One  would like  an  {\it integral  linear
  logic}, which would again treat integration as an inference rule. It
should not be  a surprise at this point  that convenient vector spaces
are extremely  well-behaved with respect to  integration. The category
\Con\ will  likely provide an  excellent indicator of  the appropriate
structure.

One can  also ask about other  classes of functions  beside the smooth
ones. Chapter 3 of \cite{KM} is devoted to the calculus of holomorphic
and  real-analytic functions  on convenient  vector spaces.  It  is an
important question as  to whether there is an  analogous comonad to be
found,  inducing the  category of  holomorphic maps  as  its coKleisli
category. Then one can  investigate whether the corresponding logic is
in any way changed.

Of course, once one has a  good notion of structured vector spaces, it
is always a good question to  ask whether one can build manifolds from
such  spaces.  Manifolds based  on  convenient  vector  spaces is  the
subject of  the latter  half of \cite{KM},  and it seems  an excellent
idea to  view these structures from the  logical perspective developed
here.

Finally, we find  categories of bornological spaces to  also be worthy
of further study from  the categorical/logical perspective. We note in
particular   the   {\it   analytic   cyclic   cohomology}   of   Meyer
\cite{Meyer}.   This  is   a  cohomology   theory  based   on  convex,
bornological vector spaces, and  provides an approach to analyzing the
{\it entire cyclic cohomology} of Connes \cite{Con}.

\bibliographystyle{plain}

\end{document}